\documentclass[modern]{aastex631}

\begin{document}

\title{Prospects for the Crossing by Earth of Comet C/2023 A3 Tsuchinshan-ATLAS's Ion Tail}

\correspondingauthor{Samuel R. Grant}
\email{samuel.grant.20@ucl.ac.uk}

\author[0000-0000-0000-0000]{Samuel R. Grant}
\affiliation{Mullard Space Science Laboratory\\ University College London\\ Holmbury St. Mary\\ Dorking, Surrey RH5 6NT, UK}
\affiliation{Centre for Planetary Sciences at UCL/Birkbeck\\ London, UK}

\author[0000-0002-5859-1136]{Geraint H. Jones}
\affiliation{European Space Agency, European Space Research and Technology Centre\\ Keplerlaan 1, 2201 AZ Noordwijk, The Netherlands}

\begin{abstract}
%<150 words
The Earth will pass approximately downstream of the previous position of comet C/2023 A3 (Tsuchinshan-ATLAS) during 2024 October 10-13. We predict that spacecraft at the Sun-Earth Lagrange Point 1, L1, have a significant likelihood to detect pickup ions from the comet, as well as changes in the solar wind associated with the crossing of the comet's ion tail. Given the Earth's magnetosphere is also likely to cross the ion tail, it is possible that geomagnetic signatures associated with this will be observed by spacecraft within the magnetosphere and possible at ground-based magnetometers, as observed during Comet 1P/Halley's apparition in 1910.
%^^^^^^^^^^^ 101 Words ^^^^^^^^^^^^^
\end{abstract}


\section*{}

On 19 May 1910, Halley's Comet, 1P/Halley, transited the Sun as observed from Earth. On the same day, multiple geomagnetic observatories across the planet measured magnetic field disturbances of varying strength. It has been suggested that this was caused by the Earth crossing the ion tail of comet Halley \citep{russell1988geomagnetic}. 

The geomagnetic disturbances measured on 19 May 1910 could be considered the earliest known in-situ measurements of a cometary ion tail, long before the nature and effects of comets were well understood. Additionally, the 1910 encounter remains the only crossing of comet Halley's tail; during Halley's 1986 perihelion, all observations were made upstream of the nucleus.

Serendipitous ion tail crossings by spacecraft are fairly infrequent, although have occurred on several occasions, e.g.  \citet{jones2000, neugebauer07}. However, although Earth should be just as likely to cross a cometary ion tail as any other object at 1~AU from the Sun, such crossings have been rare. For such a traversal to take place, a comet needs to cross the Sun-Earth line shortly before the planet is in that position, on a timeframe such that the solar wind may transport ions at several hundred km s$^{-1}$ from the comet's coma to the Earth. 

The Tailcatcher program (\emph{manuscript in prep.}) was developed to quantify the degree of alignment between the ion tail of a comet and a spacecraft or planet. This is done by the calculation of an impact parameter, that is minimised when the alignment is optimised.  
Solar wind velocity measurements made by a spacecraft anti-sunward of a comet can be used to extrapolate the flow of the solar wind back toward the Sun. The minimum distance between these extrapolated solar wind packets and known comets is a measure of the likelihood of an ion tail crossing. This method is in agreement with observations for all past identified ion tail crossings, and has twice been used to predict ion tail encounters for two comets with the spacecraft \emph{Solar Orbiter} \citep{jones2020RNAAS...4...62J, grant2022prediction}, using estimated solar wind velocities. 
Observations of the former crossing were reported by \citet{matteini2021}. Manuscripts are in preparation on the latter event.

C/2023 A3 (Tsuchinshan-ATLAS) is a 
retrograde long-period comet first identified on January 9, 2023, by two independent survey teams: the Tsuchinshan Observatory in China and the ATLAS observatory based in Hawaii. The comet reached perihelion on 24 September 2024, at a heliocentric distance of 0.391AU, before its sunward closest approach to Earth on 13 October 2024. 

\begin{figure}[h!]
\centering
\plottwo{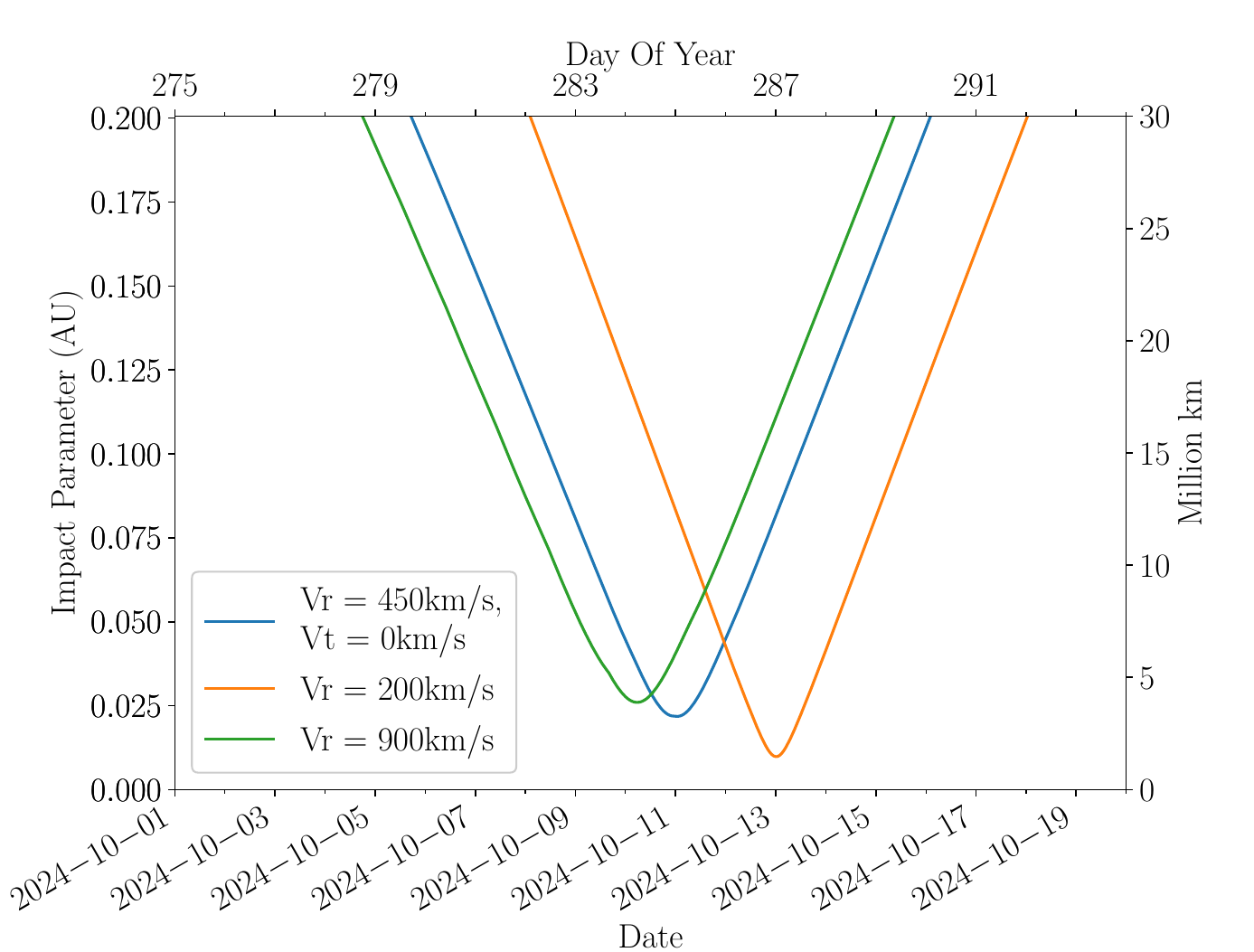}{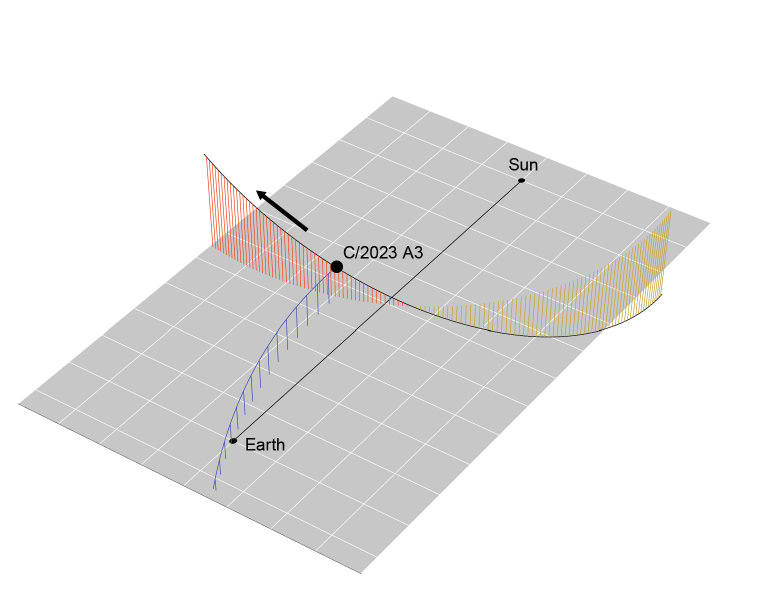}
    \caption{\emph{(Left)} Impact parameters plotted for Earth and comet C/2023 A3 (Tsuchinshan-ATLAS) for purely radial solar wind velocities 200, 450, and 900 km/s. In this case, slower solar wind speeds appear to increase the likelihood of a tail encounter. \emph{(Right)} Relative positions of the Earth, Sun, and C/2023 A3, in a frame where the Sun-Earth line is fixed. The plane represents that of the ecliptic, with grid lines spaced at 0.1~AU. The comet's position is shown every 6 hours, with legs showing its 3D position relative to the ecliptic plane - orange when south of the ecliptic, red when north. Also shown in blue is the ion tail profile simulated for 2024 October 13 at 10:00 UTC, for a radial solar wind flow of 300~km$^{-1}$. Under these circumstances, the ion tail axis would pass slightly north of Earth.} \label{tsuchinchan}
\end{figure}

As shown in Fig. \ref{tsuchinchan}, there is a fair chance that Earth will be immersed within the ion tail of the comet. Should the comet maintain a significant level of activity, geomagnetic features may be observed similarly to the comet Halley's apparition in 1910. Additionally, pickup ions from the comet will likely be intercepted by several near-Earth plasma observatories, such as NASA's \emph{Advanced Composition Explorer, ACE}.

Fig. \ref{tsuchinchan} displays the miss distances calculated via the Tailcatcher method for the Earth and comet C/2023 A3 (Tsuchinshan-ATLAS), for 3 different radial solar wind speeds. 
Assuming a perfectly radial solar wind, the slower the solar wind speed, the smaller the predicted minimum distance of the Earth from the axis of the ion tail. In reality, the solar wind often has significant non-radial flow components. This can lead to closer or further minimum impact parameters relative to the tail core; this can be assessed after the tail crossing with measured solar wind flow vectors. Note that the calculations presented here are for the main axis of the ion tail; as the tails of high-activity comets can be wide, an encounter with a flank of the tail is still possible even if the tail axis is not crossed. The neutral hydrogen corona that surrounds a comet's coma can measure several million km across, and in some cases, tens of millions of km \citep{JONES2022115199}.

As of September 24, using the visual magnitude correlation found in \citet{jorda2008correlation}, we estimate the water production rate to be of the order $Q[H_{2}O]  \sim10^{29}s^{-1}$, which places it at the same general activity level as 1P/Halley. At the time of writing, the comet's activity level has remained high, past perihelion on 27 September. Should the comet's production rate being very high, the ion tail could potentially be observable on the night side of the Earth when the comet nucleus is at low solar elongation. However, such an extended and low surface brightness feature could be challenging to detect.
The  calculated minimum miss distance for solar wind 'packets' is $\sim$(1.47-3.89)$\times$10$^{6}$~km for dates October 10-13, closer to Earth than Sun-Earth Lagrange point L1. These distances assume a purely radial solar wind flow; if the wind deviates in flow direction, as is commonly observed, then the actual miss distances could differ from these predictions.

In addition to the presence of pick-up ions, other potential solar wind signatures related to the tail crossing include a drop in solar wind speed and the structuring of the heliospheric magnetic field into a draped, dual- or multiple lobe induced magnetotail.
If the very centre of the ion tail is encountered, then a change in the charge state of solar wind ions may be observed, together with a drop in proton number density, all due to charge-exchange between solar wind ions and cometary neutral gas. 
Due to the presence of multiple near-Earth spacecraft in the solar wind, the period around 2024 October 12 therefore provides a potential opportunity for multi-point observations inside a comet tail.

Earth will cross the comet's orbital plane on 14 October at 16:10~UTC. Dust from the comet could in principle be detected around this time, but such grains would need to have been accelerated to high speeds to reach Earth and therefore would be very small in size. An observable meteor shower is therefore not expected.

\begin{acknowledgments}
\section*{Acknowledgments}
Author SRG acknowledges the support of a UKRI Science and Technology Facilities Council (STFC) research studentship. Ephemeris data used for this work was provided via the JPL Horizons system at https://ssd.jpl.nasa.gov/?horizons ; \citet{giorgini1997}.
\end{acknowledgments}

\bibliography{rnaas}{}
\bibliographystyle{aasjournal}

\end{document}